# *In Silico* Development of Psychometric Scales: Feasibility of Representative Population Data Simulation with LLMs

Enrico Cipriani[1], Pavel Okopnyi[2], Danilo Menicucci[3], Simone grassini[4]

[1] Department of Information Engineering, University of Pisa

[2] Department of Media Studies, University of Bergen

[3] Department of Surgical, Medical and Molecular Pathology and Critical Care Medicine, University of Pisa

[4] Department of Psychosocial Science, University of Bergen

## Author Note

Enrico Cipriani: https://orcid.org/0000-0001-6690-6591

Pavel Okopnyi: https://orcid.org/0000-0001-7034-2733

Danilo Menicucci: https://orcid.org/0000-0002-5521-4108

Simone Grassini: https://orcid.org/0000-0002-4189-7585

The authors declare no conflict of interest.

Correspondence concerning this article should be addressed to Simone Grassini, P.O. Box 7807, NO-5020 Bergen, Norway. Email: simone.grassini@uib.no

**Abstract**

The development and validation of psychometric scales is a resource-intensive process requiring large samples, multiple validation phases, and substantial time and funding. Recent advances in Large Language Models (LLMs) offer the possibility of generating synthetic participant data by prompting the LLM to answer items while impersonating a subject of a specific age, sex and ethnicity, potentially enabling in silico piloting of psychometric instruments before real-world data collection. Across four preregistered studies (N ≈ 300 per study), we examined whether LLM-simulated datasets can replicate the latent structures, measurement properties, and statistical characteristics of human responses. In Studies 1 and 2, we compared LLM-generated datasets to two existing validated scales using corresponding real-world samples. In Studies 3 and 4, we developed entirely new scales using EFA on simulated data, then tested whether these LLM-derived structures generalized to newly collected human samples. Results showed that simulated datasets replicated the intended factor structures in three of four studies and demonstrated consistent configural and metric invariance. Scalar invariance was achieved in the two ex-novo scales. Correlation-based tests indicated substantial differences between real and synthetic datasets, Furthermore, pronounced discrepancies emerged in score distributions and variances, indicating that while LLMs reproduce group-level latent structures, they do not approximate individual-level data properties. Simulated datasets also displayed full internal measurement invariance across gender. Overall, our findings indicate that LLM-generated data reliably support early-stage, group-level psychometric prototyping of factors and structure, but are not suitable as substitutes for individual-level validation. We discuss methodological limitations, risks of bias and data pollution, and ethical considerations surrounding the potential misuse of in silico psychometric simulations.

## Introduction

Psychometric scales and questionnaires are crucial tools of quantitative research in several fields such as psychology, medicine and health sciences, social sciences, education and human-computer interaction studies. After a novel construct or theoretical framework is proposed and becomes the subject of empirical research, the development of corresponding psychometric tools is usually a necessity. However, this process is resource-intensive and requires careful planning, time, materials, and substantial funding (Stefana et al., 2025). In this article, we will explore the possibility of reducing the resource demand of the scale development process by *in silico* prototyping psychometric scales using simulated psychometric data of representative samples generated via Large Language Models (LLMs). In the following sections, we will first describe the rationale and the hypotheses behind this research, followed by a description of four studies devised to empirically test the feasibility of this method. Finally, we will discuss our results and give recommendations for future research.

The amount of resources required for validation can stifle psychometric scale development, especially in rapidly developing research contexts where time constraints abound or where funding is a concern. For example, in academic institutions located in low-income countries. Even validating a relatively simple Likert-type scale is a process that often takes more than a year, including assessment of content validity, data collection, analysis, and dissemination of results. Validation studies necessitate large study samples: guidelines often recommend a ratio of 10 participants tested per 1 draft scale item, or a minimum of ≈ 300 participants to allow for reliable Exploratory Factor Analysis (EFA) results (Boateng et al., 2018). This number rapidly doubles when researchers want to test the test-retest reliability of the scale by administering it a second time to an independent sample. Recruiting a representative sample of participants is costly and time-consuming, pushing researchers to often rely on convenience samples of volunteering university students, which greatly limits the external validity of the resulting scale (Hanel & Vione, 2016; Wild et al., 2022). Recently, psychometric validation studies have come to rely on participant recruitment via online platforms, such as Prolific.co and Amazon MTurk (Buhrmester et al., 2018; Douglas et al., 2023; Huang et al., 2015). Cogent to the availability of funds to pay participants, these platforms have greatly reduced time expenses and may have even improved external validity by allowing the recruitment of representative samples for certain nationalities (e.g. Prolific allows to recruit representative samples from the UK and US). However, online studies performed by recruiting participants from these platforms have raised concerns for their ethical implications, as participants may come from disadvantaged backgrounds (Huang et al., 2015; Huff & Tingley, 2015; Moss et al., 2023). Also, when an exploratory study examines a construct for which no validated scale exists (often due to the novelty of the construct itself), or it is unfeasible to conduct a validation study, often researchers rely on *ad hoc* questionnaires. The construction of these scales is informed by theory only and allows researchers to largely bypass the costs of developing a scale. This, however, comes at the expense of the validity of the findings, which often fail to generalize to populations outside the study sample (M. K. Lee et al., 2020).

Another important issue in psychometric scale development is constituted by the "Jingle-Jangle Fallacy" (Gonzalez et al., 2021), that is, when two measures with the same label assess different constructs (*jingle* fallacy), or when two measures with different labels assess the same construct (*jangle* fallacy). In the latter case, due to the nuanced nature of psychological phenomena and the unclear definitions of constructs and their terminology, seemingly distinct scales can overlap dramatically, displaying high correlation coefficients. Avoiding jingle-jangle fallacies is largely a work of conceptual engineering (Chalmers, 2025) that requires a careful analysis of construct validity, criterion validity, and discriminant validity (Campbell & Fiske, 1959; Cronbach & Meehl, 1955). This entails iterating on scales via pilot testing, which can result in additional resources spent and loss of valuable research time. The lack of a systematic intervention on this phenomenon has over time resulted in the proliferation of several effectively redundant measuring instruments, which in turn resulted in a substantial fragmentation of quantitative psychological research (Anvari et al., 2025; Elson et al., 2023).

Other fields of research (e.g., pharmacology) have progressively increased the resource efficiency of their research methods by performing tests using increasingly sophisticated computer models (Colquitt et al., 2011; Shaker et al., 2021). These *in silico*[1] research methods significantly cut down on material resources and time required for traditional methods, as well as improve on ethical aspects of research by avoiding the unnecessary use of animal models or human participants (Jean-Quartier et al., 2018).

---

[1] Latin for "in the silicon": referring to elemental silicon used in computer chip manufacturing.

Likewise, the recent appearance of Generative Artificial Intelligence software (GAI) based on LLMs, such as Open AI's "Chat GPT," may pave the way for the development of *in silico* methods for fields in which computational modelling has not been seen as effective, such as social psychology. One possible application of LLMs lies in assisting researchers in the development of psychometric instruments (Beghetto et al., 2025; Hussain et al., 2024). For example, LLMs can quickly generate a pool of draft scale items to be then validated (Götz et al., 2024), through a process of "Automatic Item Generation" (AIG) (P. Lee et al., 2023, 2025). Since their architecture is based in Natural Language Processing (NLP), and they are trained using massive amounts of textual language data, LLMs are uniquely suited to address issues in scale development that originate from lexicon and semantics. For example, an LLM can quickly correct and suggest alternatives to ambiguously worded items, which can significantly impact the way participants respond by introducing errors and distortions (Tourangeau et al., 2000) and can correctly identify associations between items and their corresponding scales (Hommel & Arslan, 2024). LLMs can also aid researchers in avoiding *jingle-jangle* fallacies by using the *semantic embeddings* of scale items. With these, researchers can place scales in semantically codified spaces and observe how they cluster with each other (Rosenbusch et al., 2020; Wulff & Mata, 2025b). Scales too semantically close to each other may be redundant and may require a substantial reorganization (Wulff & Mata, 2025a). Moreover, LLMs can increase accessibility and coherence by ensuring that all items are worded at the same reading level. For an overview and recommendation on the use of LLMs in psychometric scale development, we direct the reader to a recent article by Beghetto and colleagues (2025).

Besides the generation of items and their refinement, LLMs have also recently opened another innovative possibility: the generation of simulated databases of psychometric data. LLMs can mimic human cognitive abilities (Binz & Schulz, 2023) and can simulate real-world social dynamics when interacting between themselves (Park et al., 2023). Furthermore, evidence shows that LLMs can replicate tendencies and biases present in real-world populations when prompted to impersonate specific demographics (Argyle et al., 2023; Dillion et al., 2023; Schramowski et al., 2022). LLM-simulated data has already demonstrated high similarity with real-world data: in a recent preprint manuscript by Hewitt and colleagues (2024), LLM-simulated psychometric data of a representative US sample shows strong correlations ($r \geq 0.85$) with actual experimental data from both published and unpublished studies (which the LLM could not have accessed during training). Besides commercial LLMs, new baseline models have been developed for the express purpose of generating simulated data. For example, the "Centaur" model, developed by Binz and colleagues (Binz et al., 2025), has been trained on an impressively large dataset of psychometric data. It can predict results in new samples better than existing cognitive models (cognitive models, negative log-likelihood, 0.56; one-sided t-test: $t(1,985,732) = -127.58$, $p \leq .0001$; Cohen's $d = 0.18$), and it can be generalised to new conditions (Binz et al., 2025). It is important to note that most of these studies performed analyses based on the aggregation of simulated population data, treating variability in LLM outputs as if it was actual variability in a human population. While they show results that are consistent with population norms, there is little evidence that simulated individualized data can model latent constructs in individuals (Petrov et al., 2024; Cummins, 2025). However, it is worth noting that in another preprint by Park and colleagues (2024), LLM agents can generate highly accurate (85%) simulated individual psychometric data after being fine-tuned on 2-hour individual interview transcripts. This may lead to even more precise individual-level simulated data.

  Building on this promising body of research, we explore the possibility of simulating data using LLMs to aid in psychometric scale development. This approach has the potential to drastically reduce the resources needed for the iterative piloting process by first testing scales on simulated populations. This may render scale development affordable to a broader range of psychometric researchers and may reduce the proliferation of redundant scales. In this article, we specifically tested how well state-of-the-art LLMs simulated data model factorial structures used in scale development. In the following section, we provide an outline of our research, as well as a rationale behind our research hypotheses.

## Research Hypotheses and Overview

For scale development, we largely followed the procedures proposed in the Classic Test Theory (CTT) and the recommendations for scale development outlined by Boateng and colleagues (2018). As mentioned above, this article is focused on answering one main research question (RQ):

- *RQ: "Can LLMs simulate real human responses for the purpose of validating a psychometric questionnaire?"*.

To address it, we devised a series of four preregistered studies: Study 1 compared LLM-generated data with an *in-house* dataset (i.e., collected by our own research group) that was previously used for the development of a psychometric scale. To exclude the possibility of bias due to the use of our own data, Study 2 performed the same operation but compared simulated data with a publicly available dataset gathered by another independent research group. Study 3 and Study 4 tested the predictive ability of our procedure by performing a reverse operation: instead of comparing synthetic data to already existing real-world datasets, we first developed an entirely novel psychometric scale using simulated data, and then we tested whether real-world data gathered afterwards validated the structure of the scale. Using newly gathered data allows us to pragmatically test the viability of LLM-generated data for pilot testing, as well as observe whether the LLM can generalize enough to information not included in the training set. Since a large share of datasets from the internet have been crawled on, it is an eventuality not to be excluded. These two studies differ only for content of the scale and follow the same procedure. All four studies share the same six operationalized hypotheses, which are described below:

- *H1: Responses obtained from the in silico simulated dataset align with the factor structure from the real-world dataset.*

To be usable for simulated research, LLM-generated data need to replicate latent factorial structures that are present in real-world samples. For Study 1 and Study 2, we tested this by performing a Confirmatory Factor Analysis (CFA) on simulated data to verify whether the theoretical factor structure of the scale fits the simulated data. In Study 3 and Study 4, we first hypothesized a scale factor structure by performing an Exploratory Factor Analysis (EFA) on a sample of simulated data, and we then tested whether this structure was validated by the real-world data.

- *H2: Responses obtained from the real-world dataset and those obtained from the simulated dataset possess measurement invariance.*

To be directly statistically comparable in experiments, independently gathered study samples need to possess dimensional invariance. To examine this, we performed a test of dimensional invariance by using Multigroup-CFA for using our two samples as independent groups (i.e., *real-world* and *simulated*) while introducing increasing levels of constraint. This hypothesis is broken down in tests for configural invariance (*H2.1*), metric invariance (*H2.2*), scalar invariance (*H2.3*), and residual invariance (*H2.4*). If configural invariance is confirmed, the pattern of factor loadings is equal across the group. If metric invariance is confirmed, it is also possible to perform a comparison of variances between latent variables. If scalar invariance is verified, then it is also possible to compare latent scale means. Finally, if residual invariance is confirmed, then further comparisons of manifest means and analyses with manifest scale scores are allowed (Putnick & Bornstein, 2016; Vandenberg & Lance, 2000). The achievement of high degrees of measurement invariance between real-world and simulated samples would entail that the two types of data are highly compatible, and results from *in silico* studies would reproduce those from real-world contexts.

- *H3: There is substantial agreement between responses obtained from the real-world dataset and those obtained from the simulated dataset. (H3.1: Strong correlation coefficient, H3.2: ICC)*

Hypotheses H3, H4, and H5, are less concerned with testing the dimensionality of the data and more with comparing the individual-level properties. Originally, withH£ we wanted to test whether the scores from the two groups correlate with each other. The intention was to first perform a *Spearman's* correlation test (*H3.1*) and then to compute the Intraclass Correlation Coefficient (ICC; *H3.2*). However, due to practical limitations, we could not obtain precise couples of real and simulated data for all studies except Study 2, therefore excluding pairwise correlation tests. In Study 1 we mistakenly did not generate the simulated dataset with a unique id matching each simulated participant with its real-world counterpart, therefore, we could not pair them exactly. In Study 3 and Study 4, we could not perform an exact matching due to their experimental nature: the real-world sample is recruited and collected after the simulated sample has been generated. Prolific's representative sample recruitment guarantees that participants fall into a specific age bracket but not necessarily possess a specific age. Therefore, we could not request samples exactly matching our simulated data. To obviate these limitations, we estimated correspondence between the real and simulated samples using 5000 stratified paired bootstrap resamples (stratified by age bracket, gender, and ethnicity), computing Spearman correlations in each iteration and deriving percentile-based 95% confidence intervals. Following an extensive discussion after we completed data collection, we determined that this experimental hypothesis was poorly specified, since high correlations between these datasets would only arise if questionnaire scores were dependent on the demographic variables used for data generation (i.e., gender, age, and ethnicity). Nevertheless, we tested this hypothesis, and we report its results in the spirit of preregistration and transparency in line with recent calls for limitations in analytical flexibility in research with synthetic data (Cummins, 2025).

- *H4: The distributions of scores obtained from the real-world dataset and those obtained from the simulated dataset are not significantly different from each other.*

With this hypothesis, we wanted to examine whether the data distribution of simulated samples shared the same properties as real-world data. We compared LLM-generated simulated data with the gathered real-world ones first by performing a Mann-Whitney U test to gauge differences in mean. We then compared distribution properties using a Kolmogorov-Smirnoff test.

- *H5: The variances of scores obtained from the real-world dataset and those obtained from the simulated dataset are not significantly different from each other.*

Similarly to H4, we tested whether the two samples possessed similar variances. We verified this using a Levene test of equity of variances.

- *H6: When split between males and females, the dimensional structure in the simulated dataset remains invariant.*

One crucial feature that simulated data should possess to be viable for *in silico* research is that specific subgroups should be directly comparable with each other. To test this, in each study we split simulated datasets between "Male" and "Female" simulated participants. We then performed invariance testing between them. If this hypothesis is supported, it would be feasible to conduct *in silico* comparisons between specific demographics.

**General Methods**

In this section, we describe the common methodology and workflow used in all four studies when not otherwise specified in their dedicated sections. A representation of the experimental workflow of all four studies is displayed in **Figure 1.**

**Real-World Data Sampling**

All real-world participants were recruited via Prolific (Palan & Schitter, 2018), except for the sample used in Study 2, for which we refer the reader to its corresponding validation paper (Schauffel et al., 2021). For each study, we requested a representative sample of $\approx 300$ valid participants from the United Kingdom.

**Simulated Data Generation Procedure**

The simulated data were generated using GPT4o-mini (gpt-4o-mini-2024-07-18) with a knowledge cut-off in October 2023. The default model settings were used (Temperature 1.0, Top P 1.0, Frequency penalty 0.0, Presence penalty 0.0). The LLM received a procedurally generated prompt asking it to impersonate an individual with certain demographic characteristics and to respond to the psychometric scale we intended to test. This part of the prompt was then followed by the items of the scale. The prompt also included the answer key and a request to return the answers as a single string of comma-separated values to facilitate data parsing. Each prompt was sent as a separate, independent API request with no prior conversation history (i.e., system prompt only or empty context window). This ensured that responses were not influenced by previous prompts in the same session. Below we display an example of one of the prompts used to generate the simulated data:

*"Impersonate a/an [Ethnicity] [Gender] of [Age] years of age from the United Kingdom. Answer the following [Number of items]-Item questionnaire. Each item is on a Likert scale ranging from [Likert scale range]. Use the following key: [Response Key] The questionnaire is as follows: [Questionnaire Items]*
*Give the answers as a single string of comma separated values."*

Values in brackets were procedurally inserted for each prompt from a lookup table of simulated participants. Each row of this table represented a procedurally generated individual possessing an age-ethnicity-gender combination determined by the representative UK sample stratification used by Prolific. Since Prolific performs age stratification in ranges, the exact age of each simulated individual was assigned randomly within the given range. After being generated, each LLM-generated survey reply was parsed via a simple R script and stored as a row in a spreadsheet.

Three differently worded prompts were used to account for possible idiosyncratic responses. Following the procedure described by Hewitt and colleagues (2024), we adopted an ensemble approach in which scores were averaged between prompts at the item level. Each type of prompt used in the studies is reported in full in the **Supplementary Materials** to this article. We expected that the LLM model may return an invalid (i.e., not in *comma-separated values*) answer. We opted to keep these invalid answers in the database to simulate a natural workflow. These invalid answers were treated as missing values in the later analysis process and were replaced by averaging answers from the other prompts.

**Figure 1: Study workflow**

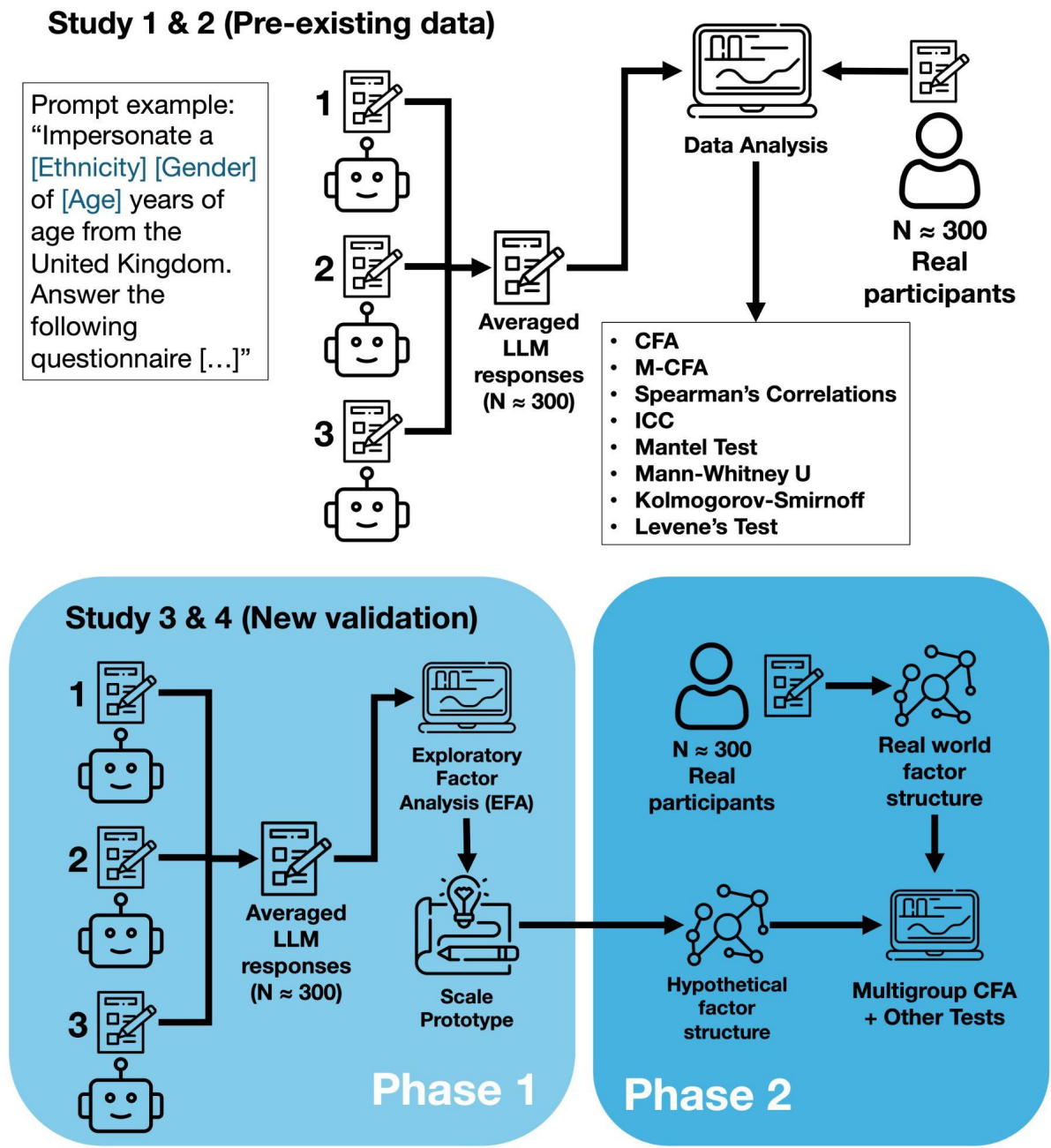

*Note.* Overview of the workflow for each study described in this article. The upper part shows the procedure for study 1 and study 2. The lower part shows the two-phase procedure for study 3 and study 4.

**Data Analysis**

To test H1, for studies 1 and 2, we performed a CFA by entering the theoretical structure of the scale of interest as a measurement model. As inferential criteria to accept or reject this hypothesis, we referred to the fit indices cut-off values proposed by Hu and Bentler (1999): CFI and TLI ≤ .900 are acceptable, CFI and TLI ≤ .950 are good. RMSEA ≤ .080 is acceptable. RMSEA ≤ .060 is good. SRMR ≤ .080 is considered acceptable.

H2 was tested by performing multigroup CFA, using the type of data as a grouping variable (simulated or real-world) and the scale structure as a measurement model. configural invariance testing was done by not introducing constraints, metric invariance testing imposed equal loadings between groups, scalar constrained loadings and intercepts, residual invariance constrained item residuals. To make inferences concerning measurement invariance, we refer to the cut-off values suggested by Chen and colleagues (Chen, 2007). A higher level of dimensional invariance is accepted only if the degree of change in CFI (ΔCFI) is equal to or less than .010, and the ΔRMSEA is equal to or less than .015.

H3.1 was tested by performing Spearman's correlation test between the real-world and simulated scale scores. H3.2 was tested by performing an ICC. H4 was tested by performing Mann-Whitney U, Kolmogorov-Smirnoff, and Levene's test of variances between the two types of data. We tested H5 by performing the same multigroup CFA procedure as H2 but on simulated data only and using gender as the grouping variable. All CFA and M-CFA models were computed using the MLR estimator due to its robustness against deviations from normality (Hox et al., 2010).

All data analysis procedures were performed using R (ver. 4.4.1) running in RStudio (ver. 2024.12.0+467). EFA, CFA, and M-CFA were performed using the "*lavaan*" package (Rosseel, 2012; ver. 0.6-19). Spearman's correlation coefficients were computed using the

"*psych*" package (Revelle, 2018; ver. 2.4.6.26). ICCs were calculated with the "*irr*" package (Gamer et al., 2012; ver. 0.84.1). Levene's test was performed via the "*car*" package (Fox & Weisberg, 2018; ver. 3.1-3).

# Study 1: Comparison Between Synthetic Data and In-house Developed Scale

## Methods

To test how different a dataset of LLM-simulated answers to a psychometric scale is from a real-world sample, our first study used the Climate Change Perceptual Awareness Scale (CCPAS) and its real-world dataset, which was previously gathered by our own research group (Cipriani et al., 2024). The CCPAS is a 15-item psychometric scale measuring individual awareness of everyday perceptual phenomena associated with climate change. It is composed of four subscales, respectively measuring awareness of climate change-linked changes in one's own affective status ("Feelings"), awareness of environmental changes related to temperature ("Temperature"), awareness of environmental changes due to droughts/humidity ("Water"), and awareness of climate-linked changes in the media discourse ("Media"). Each item is formulated as a statement to which participants respond by expressing how much they agree with it using a 5-point Likert scale. For more details on the scale and its development, we refer the reader to the original validation paper (Cipriani et al., 2024).

The hypotheses and the protocol for this study were preregistered on the OSF platform before generating the simulated data and are available at the following link: https://osf.io/km8xe/overview?view_only=e91a66c4c22e418080d358b2be160b5f

Real-world data were collected through an online survey study hosted on Microsoft Forms. The survey was launched on Friday, June 30, 2023, at 12:47 GMT, and ran until 17:30 GMT of the same day. The real-world sample (n = 316) is representative of the United Kingdom population. The simulated dataset (n = 322) was generated to match the proportions of the real-world one for three demographic variables: gender, age, and ethnicity. A full breakdown of real-world and simulated samples by demographics is displayed in **Table 1** below.

**Table 1: Demographic Breakdown of the Study Samples.**

| Var. | | Study 1 | | Study 2 | | Study 3 | | Study 4 | |
|---|---|---|---|---|---|---|---|---|---|
| | | *Real* | *Sim.* | *Real* | *Sim.* | *Real* | *Sim.* | *Real* | *Sim.* |
| **Num.** | | 316 | 322 | 331 | 331 | 301 | 300 | 301 | 300 |
| **M Age** | | 45.92 | 46.01 | 42.95 | 42.95 | 46.70 | 52.85 | 46.22 | 52.60 |
| **(SD)** | | (15.36) | (15.42) | (14.49) | (14.49) | (15.29) | (23.58) | (15.86) | (23.48) |
| **Gender** | *Men* | 154 | 156 | 153 | 153 | 146 | 144 | 146 | 144 |
| | *Women* | 162 | 166 | 178 | 178 | 154 | 156 | 152 | 156 |
| | *Other* | 0 | 0 | 0 | 0 | 1 | 0 | 3 | 0 |
| **Ethn.** | *Asian* | 20 | 22 | - | - | 23 | 23 | 23 | 23 |
| | *Black* | 9 | 10 | - | - | 10 | 10 | 10 | 10 |
| | *Mixed* | 10 | 10 | - | - | 10 | 10 | 10 | 10 |
| | *White* | 268 | 271 | - | - | 248 | 247 | 248 | 247 |
| | *Other* | 9 | 9 | - | - | 10 | 10 | 10 | 10 |

***Note.*** *"Var." = Variable. "Num." = Numerosity. "M" = Mean. "SD" = Standard Deviation. "Ethn." = Ethnicity. "Sim." = Simulated.*

**Results**

To test for H1, a CFA was performed by analyzing how the CCPAS four-factor model fit the simulated data. The results showed poor fit ($\chi^2(84) = 355.45$, $p < .001$, CFI = .827, TLI = .783, RMSEA = .100, 90% CI [.089, .111], SRMR = .102), indicating that the simulated responses did not adequately reproduce the theoretical structure (Hu & Bentler, 1999; Marsh et al., 2004). Because of this result, we reject H1.

Testing for H2, an M-CFA was conducted to assess measurement invariance of the four-factor model across the real and simulated datasets. Model fit indices and chi-square difference tests were examined sequentially for configural, metric, scalar, and residual invariance levels. The configural invariance model showed a marginal but unacceptable fit (Chen, 2007) ($\chi^2(168) = 659.18$, $p < .001$, CFI = .894, TLI = .868, RMSEA = .099, 90% CI [.091, .107], SRMR = .077), with the following models yielding progressively worse fit. The residual invariance model was outright inadmissible, as Heywood cases (i.e., negative loadings on latent factors) were present (Driel, 1978). Due to the poor fit of the configural invariance model, we reject H2.1 through H2.4. **Table 2** displays a summary of dimensional invariance results and their fit indices.

**Table 2: Dimensional Invariance Results for Study 1 (H2 Testing).**

| Model | $\chi^2$ | df | CFI | ΔCFI | RMSEA | ΔRMSEA | SRMR | Supp. |
|---|---|---|---|---|---|---|---|---|
| *Configural* (H2.1) | 659.18 | 168 | .894 | - | .099 | - | .077 | N |
| *Metric* (H2.2) | 864.92 | 179 | .853 | -.041 | .113 | .012 | .116 | N |
| *Scalar* (H2.3) | 1299.83 | 190 | .762 | -.091 | .139 | .026 | .128 | N |
| *Residual* (H2.4) | 5896.35 | 205 | .000 | -.762 | .287 | .148 | .631 | N |

*Note.* Values are robust (scaled) fit indices where available; χ² values are scaled. Decision is based primarily on *ΔCFI ≤ .010* and *ΔRMSEA ≤ .015* criteria for invariance. The column "Supp." Indicates whether there is support for the corresponding type of invariance. Y: Supported. P: Partially supported. N: Not supported.

Spearman correlations were estimated using 5,000 bootstrap resamples (percentile 95% CIs). Awareness of temperature effects showed a very small and statistically uncertain association with the outcome (ρ = .02, 95% CI [-.09, .13]). Awareness of humidity/drought effects was similarly small (ρ = .04, 95% CI [-.06, .14]). Media awareness was not associated with the outcome (ρ = .00, 95% CI [-.10, .11]). Felt experience showed a weak but unreliable association (ρ = .03, 95% CI [-.08, .13]). Given these results, we believe H3 to not be supported.

Nonparametric comparisons between real and simulated datasets revealed no significant differences in central tendency for any subscale, as indicated by Mann-Whitney U tests ($U = 46912\text{-}53698$, $p = .09\text{-}.89$). However, Kolmogorov-Smirnov tests indicated significant differences in the overall distribution of scores for all variables ($D = 0.36\text{-}0.42$, all $p < .001$). These findings suggest that although the simulated data approximated the median values of the real responses, their underlying distributions differed substantially in shape. Therefore, we believe that their results partially support H4. Histograms comparing distribution of the real and simulated dataset are displayed in **Figure 2.**

**Figure 2: Variable Distributions**

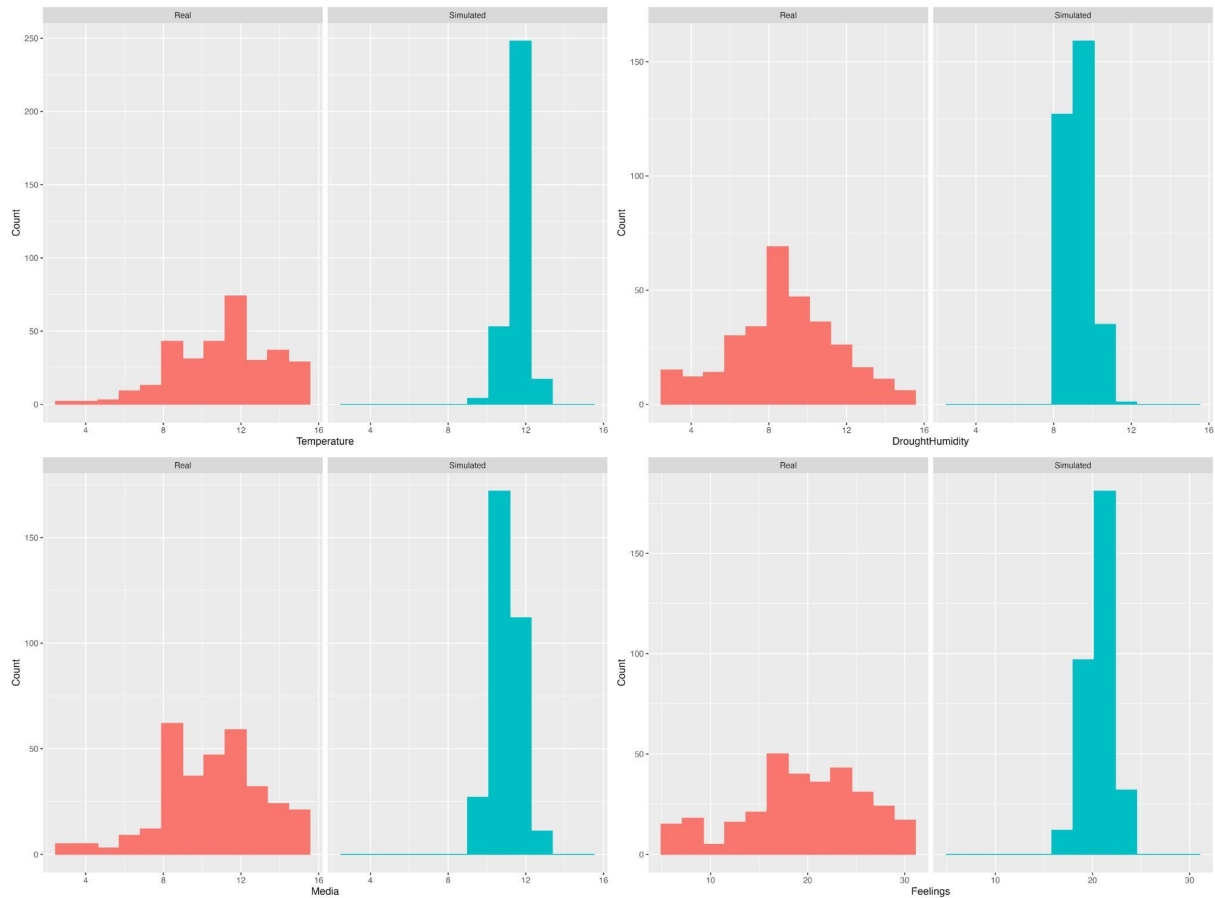

*Note. Histograms comparing the real-world sample scale distributions (red) with the simulated dataset (blue).*

Levene's tests indicated significant heterogeneity of variances across all subscales, suggesting that variability differed markedly between real and simulated datasets, $F(1, 636) = 256.16\text{-}379.85$, $p < .001$. These results confirm that the simulated responses did not reproduce the variance structure observed in the real data, and thus we reject H5.

To test H6, a series of multigroup confirmatory factor analyses explored measurement invariance of the model across gender using robust maximum likelihood estimation. The configural model showed acceptable, though not optimal, fit ($\chi^2(168) = 374.25$, $p < .001$, CFI

= .827, TLI = .784, RMSEA = .084 [.073, .096], SRMR = .097), indicating that the same factor structure held for men and women. When factor loadings were constrained to equality (metric model), fit remained comparable ($\chi^2(179) = 390.39$, $p < .001$, $\Delta$CFI = -.01), suggesting metric invariance. Constraining intercepts (scalar model) led to a substantial decline in fit ($\chi^2(190) = 609.76$, $p < .001$, CFI = .640, RMSEA = .115, $\Delta\chi^2(11) = 219.24$, $p < .001$), indicating non-invariance of intercepts. The strict model, which further constrained residuals, also fit poorly ($\chi^2(205) = 717.78$, $p < .001$, CFI = .554, RMSEA = .125, $\Delta\chi^2(15) = 99.49$, $p < .001$). Overall, the model achieved configural and partial metric invariance, but scalar and residual invariance were not supported, suggesting differences in item intercepts and residual variances between male and female respondents. Therefore, we consider H6 to be only partially supported.

## Study 2: Replication on an Independently Developed Scale

**Methods**

To exclude potential biases coming from using a dataset we previously collected ourselves, we repeated the same procedure used in Study 1 using data collected by an independent research group. For this purpose, the English dataset used to validate the "Information and Communication Self Concept, 25 Items" (ICT-SC25) was chosen. This scale measures an individual's self-perceived competence relating to information and communication technologies (Schauffel et al., 2021a). It is composed of 25 Likert-type items in total, further broken down into five subscales reflecting domain-specific aspects of ICT self-concept: *Communicate*, *Process and Store*, *Generate Content*, *Safe Application*, *Solve Problems*. Choosing this scale and its accompanying dataset was motivated by its detailed and robust multisample-multiphase validation process by the public availability of its full validation dataset (Schauffel et al., 2021b).

The scale is available both in a German-language version (ICT-SC25g) and an English one (ICT-SC25e). For our study, we used the English version and its corresponding validation sample (Sample 5, n = 483). Consistent with the other studies described in this article, this sample is a representative quota sample of the United Kingdom population. For further details on this scale and its development, we direct the reader to the validation paper by Schauffel and colleagues (2021a).

In line with Study 1, the hypotheses and procedures of this study were preregistered on OSF before generating the LLM-simulated data. The preregistration is available for consultation at the following link:

https://osf.io/w8gch/overview?view_only=87f0fe722006459d9e84ab829afb24d4

The LLM-simulated data generation followed a different procedure from Study 1 in two main aspects. First, the ICT-SC25 dataset does not contain ethnicity data; therefore our prompts to generate simulated data were modified to only contain age and gender. Second, the ICT-SC25 provides unique IDs for each participant. This allowed us to prompt simulated participants that were exactly matching real-world ones for age and gender, allowing for pairwise comparisons in H3.1 and H3.2. However, some IDs in the original dataset were duplicated; these were therefore removed to ensure exact matching of both real and simulated samples. The resulting final dataset had a numerosity of 331, still above our target of at least 300 participants. A full demographic breakdown of the real and simulated dataset is displayed in **Table 1.**

**Results**

A CFA was conducted to test the expected six-factor structure using MLR estimation. The model showed a good fit to the data ($\chi^2(260) = 975.30$, $p < .001$, CFI = .945, TLI = .936, RMSEA = .091, 90% CI [.085, .097], SRMR = .022). Overall, the model supported the proposed six-dimensional structure of the scale, therefore supporting H1.

An M-CFA was conducted to examine measurement invariance of the six-factor model across the real and simulated datasets. The configural model showed a good fit ($\chi^2(520) = 1478.03$, $p < .001$, CFI = .953, TLI = .946, RMSEA = .084, 90% CI [.079, .089], SRMR = .024), indicating that the same factor structure was appropriate across groups. When computing a metric invariance model, fit slightly decreased ($\chi^2(539) = 1850.90$, $p < .001$, CFI = .937, TLI = .929, RMSEA = .096, SRMR = .092). Scalar invariance further worsened model fit ($\chi^2(558) = 2495.51$, $p < .001$, CFI = .906, TLI = .899, RMSEA = .115, SRMR = .098). The strict invariance model showed poor fit ($\chi^2(583) = 4427.52$, $p < .001$, CFI

= .819, TLI = .814, RMSEA = .156, SRMR = .134). Together, these results indicate configural invariance (H2.1) and partial support for metric invariance, due to ΔCFI superating the acceptability threshold (H2.2), but not scalar or strict invariance (H2.3, H2.4). **Table 3** displays a comparison of fit indices between models.

**Table 3: Dimensional Invariance Results for Study 2 (H2 Testing).**

| Model | $\chi^2$ | df | CFI | ΔCFI | RMSEA | ΔRMSEA | SRMR | Supp. |
|---|---|---|---|---|---|---|---|---|
| *Configural (H2.1)* | 1478.032 | 520 | .953 | - | .084 | - | .024 | Y |
| *Metric (H2.2)* | 1850.897 | 539 | .937 | -.016 | .096 | .012 | .092 | P |
| *Scalar (H2.3)* | 2495.508 | 558 | .906 | -.031 | .115 | .019 | .098 | N |
| *Residual (H2.4)* | 4427.519 | 583 | .819 | -.087 | .156 | .041 | .134 | N |

*Note.* Values are robust (scaled) fit indices where available; $\chi^2$ values are scaled. Decision is based primarily on *ΔCFI* ≤ .010 and *ΔRMSEA* ≤ .015 criteria for invariance. The column "Supp." Indicates whether there is support for the corresponding type of invariance. Y: Supported. P: Partially supported. N: Not supported.

Testing for H3.1, cross-dataset Spearman's correlations between matching dimensions were moderate (e.g., ρ ranging .21-.35, p < .001), suggesting that while the real and simulated responses reflected comparable patterns, they were not interchangeable. **Table 4** below displays a matrix of correlations between the real and simulated ICT-SC25 subscales. To further assess consistency between real and simulated scores (H3.2), a two-way, single measure intraclass correlation coefficient (i.e., ICC(A,1)) was computed, yielding ICC = .19, 95% CI [.09, .29], $F(330, 275) = 1.52$, $p < .001$. This indicates significant but poor absolute agreement, implying that although dimensional structures were similar, individual-level scores differed substantially between the two datasets. Given these results, H3 was only partially supported.

**Table 4: Correlation Matrix (Spearman's ρ) Results for Study 2 (H3.1 Testing).**

| Variable | 1. (Real) | 2. (Real) | 3. (Real) | 4. (Real) | 5. (Real) | 6. (Real) |
|---|---|---|---|---|---|---|
| *1. General (Simulated)* | **0.27*** | 0.29*** | 0.24*** | 0.24*** | 0.24*** | 0.35*** |
| *2. Communicate (Simulated)* | 0.28*** | **0.3*** | 0.24*** | 0.26*** | 0.24*** | 0.35*** |
| *3. Process Store (Simulated)* | 0.26*** | 0.29*** | **0.24*** | 0.24*** | 0.23*** | 0.33*** |
| *4. Generate Content (Simulated)* | 0.26*** | 0.28*** | 0.23*** | **0.24*** | 0.21*** | 0.34*** |
| *5. Safe Application (Simulated)* | 0.25*** | 0.24*** | 0.19*** | 0.19*** | **0.2*** | 0.29*** |
| *6. Solve Problems (Simulated)* | 0.29*** | 0.29*** | 0.24*** | 0.25*** | 0.24*** | **0.34*** |

*Note.* Non-parametric pairwise correlations between real and simulated datasets. Coefficients in the diagonal (in bold) represent associations between variables of the simulated dataset and their real-world counterpart. All coefficients are Spearman's ρ. *** = $p < .001$.

Mann-Whitney U tests indicated significant group differences for all domains: General (U = 83 365, $p < .001$), Communicate (U = 76 511, $p < .001$), Process & Store (U = 71 270, $p < .001$), Generate Content (U = 60 784, $p = .015$), Safe Application (U = 63 947, $p < .001$), and Solve Problems (U = 62 426, $p = .002$). Complementary two-sample

Kolmogorov-Smirnov tests confirmed these results, showing significant distributional divergence between real and simulated data across all constructs (all Ds = 0.34-0.57, all ps < .001). Taking these together, we reject H4. Histograms comparing distribution of the real and simulated dataset are displayed in **Figure 3.**

**Figure 3: Variable Distributions**

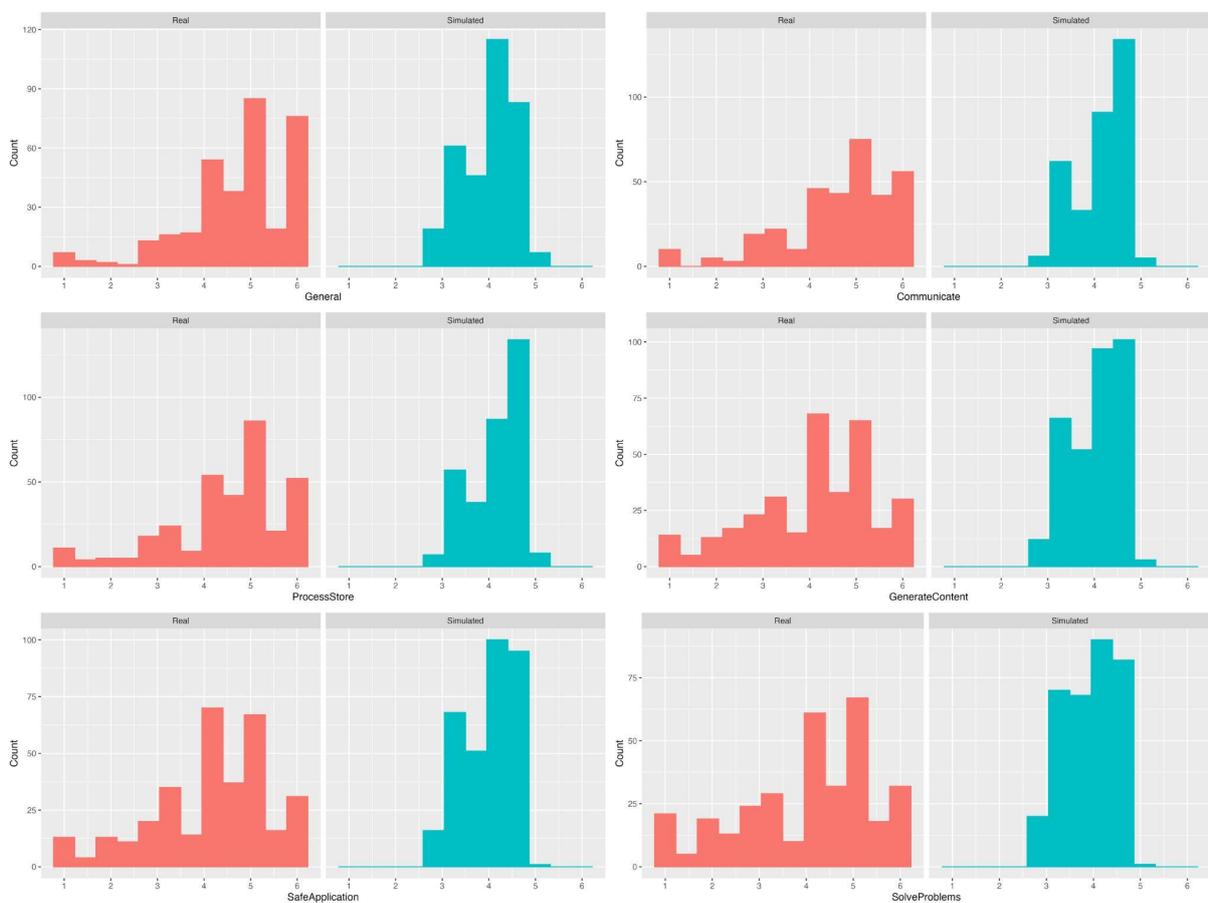

*Note.* Histograms comparing the real-world sample scale distributions (red) with the simulated dataset (blue).

Levene's test results indicated significant heterogeneity of variance for every domain (All p < .001). These findings indicate that the assumption of homogeneity of variance was violated; therefore, H5 is not supported.

An M-CFA was conducted to examine measurement invariance across gender in the simulated data. The configural model demonstrated good fit ($\chi^2(520) = 1114.49$, $p < .001$, CFI = .954, TLI = .947, RMSEA = .083, 90% CI [.076, .090], SRMR = .021), indicating that the same factor structure was appropriate for both men and women. Testing for metric invariance produced minimal changes in fit ($\chi^2(539) = 1151.36$, $p < .001$, CFI = .953), supporting metric invariance. Model fit for scalar invariance was acceptable but a substantial decrease compared to the previous ($\chi^2(558) = 1423.97$, $p < .001$, CFI = .932). The residual invariance model yielded acceptable fit ($\chi^2(583) = 1455.27$, $p < .001$, CFI = .931). Overall, these results support configural and metric invariance, with limited evidence for scalar invariance and acceptable residual invariance across male and female simulated data. Considering this, we deem H6 to be supported.

## Study 3: Replication on an Ex-novo Scale

**Methods**

After testing for LLM capabilities to simulate already existing scales in Study 1 and Study 2, LLM's predictive capacity to develop a novel scale was evaluated. We set out to develop the "Shame And Guilt for AI Test" (SAGAT): a novel scale for the measurement of feelings of Shame, Guilt, and Impostor syndrome linked to the use of AI tools. The initial development of the scale followed the guidelines and recommendations described by Boateng and colleagues (Boateng et al., 2018): after generating a pool of draft items, they were submitted to a panel of experts, which rated each item for its content validity. Following this, only the items with high Content Validity Indices and inter-rater agreement were retained (CVIs, Lynn, 1986). The retained draft items were then submitted to the LLM following the same procedure used to generate simulated data in Studies 1 and 2. The simulated data were then used to perform iterations of exploratory factor analysis to create a *scale prototype*. This *in silico* prototyping process we used to develop this scale using our simulated data is described in detail in the **Supplementary Materials** to this article, as well as in the preregistration. As for the other studies, the preregistration and its accompanying documents are available on OSF at the following link:

https://osf.io/h6aj7/overview?view_only=34b37bdfa1ff402db266fb18df6478e3

The prototyped scale is composed of 9 items on a 7-point Likert scale, broken down into 3 subscales measuring AI use-linked feelings of *Shame, Guilt,* and *Impostor Syndrome,* respectively. The real-world dataset was obtained by performing a survey study that contained the after-CVI retained draft scale items, as well as other psychometric scales to test the criterion validity of our scale. The online survey ran on Prolific from Monday, 03 February 2025, at 18:00 GMT, to Tuesday, 04 February 2025, at 23:30 GMT.

Once we obtained the real-world dataset, we proceeded to compare it with the previously generated simulated dataset, using the same data analysis procedure as Study 1. A demographic comparison of the two datasets is displayed in **Table 1**. The SAGAT scale will be described in more detail in a future scientific article, after completing a traditional scale development process partially informed by our experimental prototyping method.

**Results**

To assess H1, we performed a CFA on the real-world dataset to evaluate how well the three-factor structure of the SAGAT we extracted from simulated data will fit the real data. The model demonstrated a good fit to the data ($\chi^2(24) = 60.50$, $p < .001$, CFI = .963, TLI = .944, RMSEA = .080, 90% CI [.055, .106], SRMR = .047). All standardized factor loadings were significant ($p < .001$) and ranged from .65 to .89, indicating that the indicators loaded strongly on their intended factors. The three latent constructs were moderately to strongly correlated (*Impostor-Shame* = .76, *Impostor-Guilt* = .79, *Shame-Guilt* = .63). This result suggests that the theoretical structure developed from simulated data is replicated in a real-world dataset, thus supporting H1.

Measurement invariance of the AI Shame Scale was examined across the real and simulated datasets using M-CFA. The configural model showed good fit to the data ($\chi^2(48) = 97.86$, $p < .001$, CFI = .980, TLI = .970, RMSEA = .063, 90% CI [.045, .081], SRMR = .035). Imposing metric invariance produced a slightly reduced but still acceptable fit ($\chi^2(54) = 123.24$, $p < .001$, CFI = .972, TLI = .963, RMSEA = .070, 90% CI [.053, .086], SRMR = .052). The change in CFI ($\Delta$CFI = -.008) was well below the recommended .010 threshold, supporting metric invariance. The scalar model also demonstrated acceptable fit ($\chi^2(60)$ = 159.30, $p < .001$, CFI = .961, TLI = .953, RMSEA = .079, 90% CI [.064, .094], SRMR = .058). The small decrease in fit relative to the metric model ($\Delta$CFI = -.011) suggests approximate scalar invariance. Finally, the residual invariance model showed a severe deterioration in fit ($\chi^2(69)$ = 4494.37, CFI = .000, TLI = -.27, RMSEA = .409, SRMR = .590), and it is inadmissible, as Heywood cases are present (Driel, 1978). **Table 5** displays a summary of the models and their changes in fit. Overall, configural (H2.1), metric (H2.2), and scalar invariance (H2.3) were supported across real and simulated data.

**Table 5: Dimensional Invariance Results for Study 3 (H2 Testing).**

| Model | $\chi^2$ | df | CFI | $\Delta$CFI | RMSEA | $\Delta$RMSEA | SRMR | Supp. |
|---|---|---|---|---|---|---|---|---|
| *Configural* (H2.1) | 97.86 | 48 | .980 | - | .063 | - | .035 | Y |
| *Metric* (H2.2) | 123.24 | 54 | .972 | -.008 | .070 | .009 | .052 | Y |
| *Scalar* (H2.3) | 159.30 | 60 | .961 | -.011 | .079 | .009 | .058 | Y |
| *Residual* (H2.4) | 4494.37 | 69 | .000 | -.961 | .409 | .330 | .590 | N |

*Note. Values are robust (scaled) fit indices where available; $\chi^2$ values are scaled. Decision is based primarily on $\Delta$CFI $\leq$ .010 and $\Delta$RMSEA $\leq$ .015 criteria for invariance. The column "Supp." Indicates whether there is support for the corresponding type of invariance. Y: Supported. P: Partially supported. N: Not supported.*

To test for H3, we estimated Spearman's correlations using 5,000 bootstrap resamples with percentile-based 95% confidence intervals. The association with impostor syndrome was very small and not statistically reliable, ρ = .01, 95% CI [-.11, .13]. Shame also showed no reliable association, ρ = .00, 95% CI [-.12, .12]. Guilt showed a weak and statistically uncertain positive association, ρ = .06, 95% CI [-.05, .18]. These results suggest no significant correspondence between the datasets; therefore we reject H3.

Group comparisons between real and simulated data were conducted using the Mann-Whitney U test. For the *Impostor Syndrome* factor, the Wilcoxon rank-sum test indicated no significant difference between real and simulated scores (U = 41,516, p = .13). In contrast, significant differences emerged for both the *Shame* (U = 29,457, p < .001) and *Guilt* (U = 26,392, p < .001) dimensions. These results were corroborated by Kolmogorov-Smirnov tests, which revealed significant distributional differences for all three constructs (all p values < .001). Overall, these findings do not support H4. Histograms comparing distribution of the real and simulated dataset are displayed in **Figure 4.**

**Figure 4: Variable Distributions**

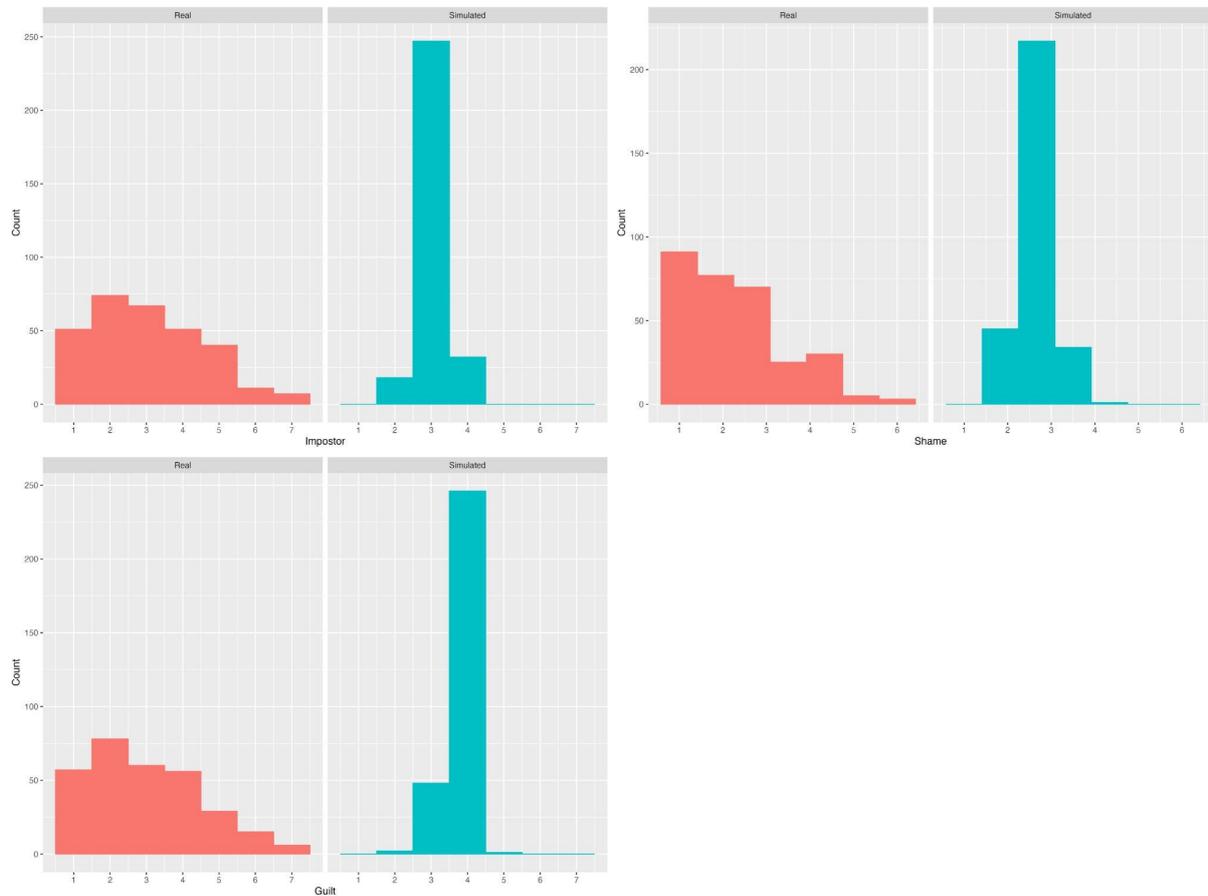

*Note. Histograms comparing the real-world sample scale distributions (red) with the simulated dataset (blue).*

Levene's tests indicated significant heterogeneity of variance between the real and simulated datasets across all three constructs (all p values < .001). These results suggest that the variability of scores differed substantially between real and simulated data across all dimensions of the SAGAT; thus, we reject H5.

Internal measurement invariance of the SAGAT was tested in the simulated dataset across gender groups using M-CFA. The configural model demonstrated an excellent fit to the data ($\chi^2(48) = 59.14$, $p = .13$, CFI = .993, TLI = .990, RMSEA = .039, 90% CI

[.000, .069], SRMR = .033). When metric invariance was imposed, model fit remained excellent ($\chi^2(54)$ = 62.70, p = .20, CFI = .995, TLI = .993, RMSEA = .033, 90% CI [.000, .063], SRMR = .034). Adding scalar invariance constraints resulted in a similarly strong fit ($\chi^2(60)$ = 69.63, p = .19, CFI = .994, TLI = .993, RMSEA = .033, 90% CI [.000, .062], SRMR = .036), supporting scalar invariance. Finally, the residual invariance model also fit well ($\chi^2(69)$ = 75.28, p = .28, CFI = .996, TLI = .996, RMSEA = .025, 90% CI [.000, .055], SRMR = .039). No meaningful loss of fit was observed, confirming full measurement invariance across gender. These results strongly support H6.

## Study 4: Second Replication on an Ex-novo Scale

**Methods**

To increase the validity and reliability of our findings, we performed a replication study by developing another novel scale. For this study, we decided to develop a scale to measure AI-linked anxious belief. This "AI Anxiety" scale was developed using the same procedure as per Study 3, which can be retrieved in its corresponding "Methods" section. For an even more detailed report of this procedure, the reader will find the *in silico* development process of the scale using simulated data is displayed in the **Supplementary Materials** to this article, as well as in the preregistration, which can be consulted on the OSF at the following link:

https://osf.io/qjcnd/overview?view_only=b32bc6344c2c44be8c23d8758cf8f0f9

The prototyped "AI Anxiety" scale is composed of 13 items on a 7-point Likert scale. These items are further divided into five subscales: "*Competence Concerns*" (worry about being able to use AI competently), "*Job Concerns*" (worry about reduced occupational opportunities due to AI), "*AI Adaptation Stress*" (feelings of stress associated with interaction with AI), "*Privacy Concerns*" (AI-linked worry about privacy loss), and "*Consciousness Concerns*" (worry related to the possibility that AI tools may become conscious).

The survey conducted to gather the real-world dataset for comparison with the simulated one ran on Prolific from Monday, 03 March 2025, at 15:00 GMT, to Tuesday, 04 March 2025, at 17:00 GMT. A demographic breakdown of the two datasets is displayed in **Table 1**. Similar to the SAGAT scale, the "AI Anxiety" scale will be submitted for publication after conducting a traditional validation study.

**Results**

A CFA was conducted to examine the proposed five-factor structure of the AI Anxiety Scale prototype. The model demonstrated an adequate fit to the real-world data ($\chi^2(55) = 175.33$, $p < .001$, CFI = .943, TLI = .920, RMSEA = .088, 90% CI [.074, .103], SRMR = .086). All standardized factor loadings were significant ($p < .001$) and generally strong ($.42 \leq \lambda \leq .93$), supporting the hypothesized latent structure. Inter-factor correlations ranged from moderate to high, indicating conceptual relatedness among the subdimensions: job concern showed positive associations with AI Adaptation Stress ($r = .55$), Privacy Concerns ($r = .66$), and Consciousness Concerns ($r = .59$). At the same time, competence concern was negatively related to Job Concerns ($r = -.36$) and AI Adaptation Stress ($r = -.33$). Together, these results support H1.

Measurement invariance of the AI Anxiety Scale was assessed across the real and simulated datasets using M-CFA. The configural model showed a good fit ($\chi^2(110) = 300.26$, $p < .001$, CFI = .962, TLI = .947, RMSEA = .078, 90% CI [.068, .089], SRMR = .065). Computing metric invariance slightly reduced model fit ($\chi^2(118) = 341.08$, $p < .001$, CFI = .956, TLI = .942, RMSEA = .082, 90% CI [.072, .092], SRMR = .068). The small change in CFI ($\Delta$CFI = -.006) supported metric invariance. Adding scalar invariance constraints resulted in a comparable fit ($\chi^2(126) = 385.62$, $p < .001$, CFI = .949, TLI = .937, RMSEA = .086, 90% CI [.076, .095], SRMR = .071). The decrease in CFI ($\Delta$CFI = -.007) indicated that scalar invariance was maintained. In contrast, the residual invariance model demonstrated a substantial loss of fit ($\chi^2(139) = 3860.87$, $p < .001$, CFI = .390, TLI = .315, RMSEA = .281, SRMR = .302), suggesting that residual variances differed significantly between real and simulated datasets. Overall, these results support configural (H2.1), metric (H2.2), and scalar (H2.3) invariance, but not residual invariance (H2.4), indicating that while the latent structure and measurement properties are largely equivalent between real and simulated data, residual variability remains higher in the simulated responses. **Table 6** below summarizes the model fit indices.

**Table 6: Dimensional Invariance Results for Study 4 (H2 Testing).**

| Model | χ² | df | CFI | ΔCFI | RMSEA | ΔRMSEA | SRMR | Supp. |
|---|---|---|---|---|---|---|---|---|
| *Configural* (H2.1) | 300.26 | 110 | .962 | - | .078 | - | .065 | Y |
| *Metric* (H2.2) | 341.08 | 118 | .956 | -.006 | .082 | .004 | .068 | Y |
| *Scalar* (H2.3) | 385.62 | 126 | .949 | -.007 | .086 | .004 | .071 | Y |
| *Residual* (H2.4) | 3860.87 | 139 | .390 | -.559 | .281 | .195 | .302 | N |

*Note.* Values are robust (scaled) fit indices where available; χ² values are scaled. Decision is based primarily on *ΔCFI ≤ .010* and *ΔRMSEA ≤ .015* criteria for invariance. The column "Supp." Indicates whether there is support for the corresponding type of invariance. Y: Supported. P: Partially supported. N: Not supported.

Spearman's correlations were computed with 5,000 bootstrap resamples (percentile 95% CIs). Competence-related concerns were not associated with the outcome (ρ = -.01, 95% CI [-.12, .11]). Job-related concerns showed no reliable relationship (ρ = .00, 95% CI [-.11, .12]). AI adaptation concerns showed a weak but uncertain association (ρ = .02, 95% CI [-.10, .14]). Privacy concerns were similarly small and uncertain (ρ = .02, 95% CI

[-.10, .14]). Consciousness-related concerns were effectively null ($\rho$ = .00, 95% CI [-.11, .12]). Considering these results, we reject H3.

Group comparisons between the real and simulated datasets were conducted using Mann-Whitney U. The tests revealed significant differences for *Competence Concerns* (U = 51302, p = .004), *AI Adaptation Stress* (U = 27453, p < .001), *Privacy Concerns* (U = 39036, p = .004), and *Consciousness Concern*s (U = 32962, p < .001), while *Job Concerns* showed no significant difference between datasets (U = 45134, p = .99). These results were corroborated by Kolmogorov-Smirnov tests (all p values < .001). Taken together, these results suggest that H4 is not supported. Histograms comparing distribution of the real and simulated dataset are displayed in **Figure 5**.

**Figure 5: Variable Distributions**

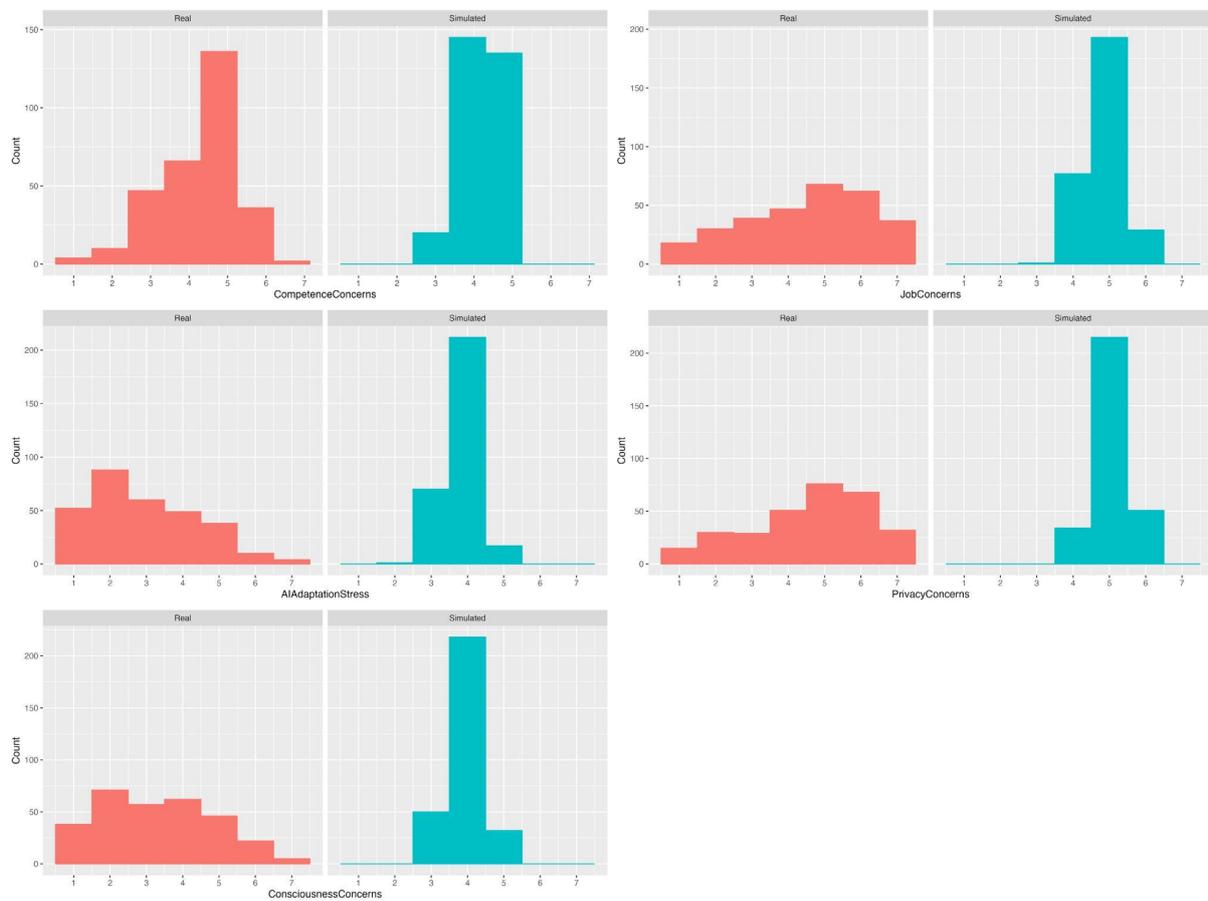

*Note.* Histograms comparing the real-world sample scale distributions (red) with the simulated dataset (blue).

Levene's tests for equality of variances indicated significant heterogeneity of variance between the real and simulated datasets across all five dimensions of the AI Anxiety Scale (all p values < 0.001). These results demonstrate that the simulated dataset exhibited substantially different variability from the real dataset across all latent dimensions, therefore rejecting H5.

Internal measurement invariance of the AI Anxiety Scale in the simulated dataset was examined across gender groups using M-CFA. The configural model demonstrated a good fit

to the data ($\chi^2(110) = 195.92$, $p < .001$, CFI = .972, TLI = .960, RMSEA = .073, 90% CI [.056, .090], SRMR = .054). When computing a metric invariance model, fit remained good and stable ($\chi^2(118) = 208.58$, $p < .001$, CFI = .970, TLI = .960, RMSEA = .073, 90% CI [.056, .088], SRMR = .062). Introducing scalar invariance produced a similar fit ($\chi^2(126) = 227.70$, $p < .001$, CFI = .967, TLI = .959, RMSEA = .074, 90% CI [.059, .090], SRMR = .063). Finally, testing for residual invariance resulted in a similar fit ($\chi^2(139) = 243.04$, $p < .001$, CFI = .965, TLI = .961, RMSEA = .072, 90% CI [.057, .087], SRMR = .065). Overall, the AI Anxiety Scale demonstrated full measurement invariance between male and female participants in the simulated dataset, confirming H6.

# General Discussion

In **Table 7** below, we display an overview of the tested hypothesis across our four studies.

**Table 7: Summary of Hypothesis Testing Across Studies**

| Hypothesis | Study 1 | Study 2 | Study 3 | Study 4 |
|---|---|---|---|---|
| *H1 (Equality of factor structures)* | Rejected | **Supported** | **Supported** | **Supported** |
| *H2.1 (Configural Invariance)* | Rejected | **Supported** | **Supported** | **Supported** |
| *H2.2 (Metric Invariance)* | Rejected | **Partially Supported** | **Supported** | **Supported** |
| *H2.3 (Scalar Invariance)* | Rejected | Rejected | **Supported** | **Supported** |
| *H2.4 (Residual Invariance)* | Rejected | Rejected | Rejected | Rejected |
| *H3 (Cross-dataset Correlations)* | Rejected | **Partially Supported** | Rejected | Rejected |
| *H4 (Equality of distributions)* | **Partially Supported** | Rejected | Rejected | Rejected |
| *H5 (Equality of variances)* | Rejected | Rejected | Rejected | Rejected |
| *H6 (Internal measurement invariance)* | **Partially Supported** | **Supported** | **Supported** | **Supported** |

Except for Study 1, all simulated datasets successfully replicated the factor structure of their corresponding scale (H1). Similarly, M-CFA shows that simulated participants and real-world ones respond to the scales in comparable ways (H2). These results are in line with previous research on group-level simulated data (Argyle et al., 2023; Hewitt et al., 2024). Remarkably, Studies 3 and 4 achieve high levels of measurement invariance between datasets in comparison to Studies 1 and 2, possibly suggesting that scales developed from simulated subjects are more generalizable to human ones than the contrary. One possible interpretation is that, when responding, LLMs tend to interpret scale items using a more standardized "common semantic denominator," less influenced by randomness and biases due to study conditions and population characteristics. Therefore, factor structures extracted from simulated samples may be more semantically generalizable to external samples than those derived from real-world data. Further research should explore factors shaping differences in how LLMs and human subjects interpret and respond to questionnaire items.

While CFA yields reliably positive results, simulated samples fail to replicate pairwise correlations (H3), the distribution (H4) and the variance (H5) of real-world samples. Taken together, these results indicate that although both datasets share the same factorial structure, their score distributions differ significantly across all measured dimensions. Therefore, group-level data and their dimensional structures are reliable enough to be used in pilot testing. However, individual-level simulated data are not suitable for this purpose. These limitations are in line with what has been observed by Petrov and colleagues (2024) and more recently by Cummins (2025).

As other researchers have been pointing out, one potential reason for which LLMs fall short in generating individual-level data is that while we are treating the synthetic data as independent observations sampled from a population, what we are doing is drawing repeatedly from the same LLM model: a single conditional probability distribution whose parameters are fixed by training. Further, we are treating "noise" in output generation due to stochastic sampling procedures (such as temperature) applied to its probability distribution, as if it was genuine real-world variability stemming from individual differences (B. May, Personal communication, December 22, 2025). This is a category error that may upend the use of traditional statistical testing on synthetic data, and for which a careful reconsideration of the current research literature comparing real-world and LLM-generated data may be necessary (B. May, Personal communication, December 22, 2025).

Specifying demographic characteristics like we did in our study may introduce something akin to individual differences. However, as recently observed by Cummins (2025), adding detailed demographic information only provides marginal improvements. Specifying sample characteristics may influence the simulation of some specific measures, but not of others: as mentioned earlier when discussing the rationale of H3, it is possible that variations in score of LLM outputs more accurately simulate variations of real-world ones only when the examined construct is associated with the participant characteristics that are specified in the prompt (in our case, the demographics). In other words, if we are simulating a psychometric score that is highly associated with age (e.g. conscientiousness; Donnellan & Lucas, 2008), the variability in simulated data may match the real-world ground truth more closely, as the LLM will use the age information we provided in the prompt to generate a matching output. Likewise, if we are simulating a psychometric variable that is not associated with age, adding this demographic information to the prompt will not significantly increase the quality of synthetic data.

It is worth noting that while Studies 2, 3, and 4 yield similar results, Study 1 shows remarkable differences. These can be imputable to several factors: first, the dimensional structure of this scale was determined using a different rotation (minres) than other scales, which may not have resulted in different responses (Sass & Schmitt, 2010). Second, this scale uses a 5-point Likert scale for responses, while others use a 6-point (Study 2), or a 7-point one (Studies 3 and 4). It has been reported in literature that changing the number of response categories in Likert scales alters their psychometric properties (Leung, 2011); this variation may have altered the LLM response pattern. Third, the subject matter of the scale (climate change perception) is a politically contextualised topic, and it is subject to massified online disinformation efforts (Lewandowsky, 2021). The body of training data on which the LLM has been trained may have been "polluted," thus causing a discrepancy between real-world and simulated data (Pan et al., 2023).

Studies performed on simulated data can have positive impacts for society, for example, in public health by testing communication strategies to improve adoption of healthy behaviour (Hewitt et al., 2024). However, it is important to highlight that this methodology can be abused for non-ethical purposes. For example, a malevolent political actor may attempt to devise and test destabilising propaganda strategies by using samples entirely made of simulated data. If the actor in question is in possession of the appropriate computational infrastructure, this *in silico* research can be done covertly using local LLMs. Psychometric scales developed this way can be used for all manners of non-ethical purposes, such as measuring which segments of the general population are more susceptible to certain destabilization techniques.

## Limitations

While robust, our research process and findings possess several notable limitations. First, we only tested Likert-type scales. Further testing should be done on other types of psychometric scales. We also have only tested survey studies: more research should be done for psychometric tools that rely on behavioural or other types of data. This work cannot fully demonstrate that the ability of LLM to generate valid synthetic data for prototyping scales is the same for all possible areas of investigation of individual psychological and behavioural characteristics.

Our studies made use of real-world English-speaking samples from the United Kingdom. This limits the generalizability of our results for other languages and other socio-cultural contexts. LLM responses to psychological testing are consistent with those of W.E.I.R.D. (i.e., Western, Educated, Industrialized, Rich, Democratic) populations (Atari et al., 2023). For this research we only used a specific ChatGPT model; therefore, further research should be done on newer or alternative LLMs to test their capabilities. Finally, the tested scales possess a relatively small number of items (the largest was Study 4, with 30 draft items tested). It is not clear how the size of the context may alter the response pattern and reliability of the LLM outputs.

## Conclusions

In a series of four robust preregistered studies with nationally representative samples, we explored how well LLM-simulated datasets replicate factor structures for the purpose of psychometric scale development. Simulated datasets reliably replicate factor structures of real-world ones, rendering them suitable for piloting early stages of scale validation processes such as EFA. However, while simulated datasets are reliable for group-level data, they are not for individual-level measurements. We further caution researchers to make critical use of simulated data, as they are also susceptible to the limitations of the LLM generating them, such as algorithmic bias and manipulation. Overall, we conclude that using LLM-simulated data may be a quick, reliable, and cost-effective procedure for *in silico* piloting of psychometric scale studies.